\input amssym.def
\documentstyle{article}

\newcommand{\bu}{1\!\rule{0.15mm}{2.1mm}
\hspace{-0.7mm}\rule[2.1mm]{1mm}{0.1mm}
\hspace{-0.5mm}\rule{0.6mm}{0.1mm}}
\title{ARBITRARY SUPERSPIN MASSIVE SUPERPARTICLES}
\author{S.M. Kuzenko, S.L. Lyakhovich\footnote{e-mail:
SLL@fftgu.tomsk.su}, and A.Yu. Segal}
\date{Department of Physics, Tomsk State University\\
634050 Tomsk, Russia}
\begin{document}
\maketitle

\stepcounter{footnote}
\begin{abstract}
We propose the action for a massive $N$-extended superparticle with
a pure (half)integer superspin $Y,~Y = 0, 1/2, 1, 3/2, \ldots$.
Regardless of the superspin value, the configuration space is ${\Bbb
R}^{4|4N} \times S^2$, where $S^2$ corresponds to spinning degrees of
freedom. Being explicitly super-Poincar\'e invariant, the model
possesses two gauge symmetries implying strong conservation of the
squared momentum and superspin. Hamilton constrained dynamics is
developed and canonical quantization is studied. For $N$ = 1 we show
that the physical super wave-functions are to be on-shell massive
chiral superfields. Central-charges generalizations of the model are given.
\end{abstract}
\newpage

Recently we have suggested [1] a new model called the ($m,s$)-particle.
This model describes a relativistic massive spin-$s$ particle ($s$
being a fixed integer or half-integer) on Minkowski space as a particle
living on ${\Bbb R}^4 \times S^2$. As was shown in Ref. [1], the principles
underlying the ($m,s$)-particle model have clear physical and geometrical
origin. The model of ($m,s$)-particle appears to be a minimal theory for
massive spinning particles in the following sense. First, its configuration
space ${\cal M}^6 = {\Bbb R}^4 \times S^2$, universal for any spin, has
minimal dimension to describe the dynamics of a particle with spin.
Second, the strong physical observables (that is, those phase-space
functions which commute with the first-class constraints) are functions
of the Hamilton generators of the Poincar\'e group only. It is remarkable
also that our model admits a number of nontrivial generalizations, in
particular, supersymmetric extensions to flat global and anti-de Sitter
superspaces. It also supplies us with new means to study interaction
problem, which still remains actual for both higher spin fields and
particles.

In the present letter we suggest the globally supersymmetric model of
a massive $N$-extended superparticle with a fixed superspin $Y$, which
will be called the model of ($m,Y$)-superparticle. The configuration
space is ${\Bbb R}^{4|4N} \times S^2$, ${\Bbb R}^{4|4N}$ being the
$N$-extended superspace parametrized by 4 even and $4N$ odd coordinates
($x^a, \theta^{\alpha I}, {\bar\theta}^{\dot\alpha}_{~I}$), where $a$ =
0, 1, 2, 3, $I = 1, \ldots, N$, and the two-component spinor indices
$\alpha,\dot\alpha$ take values 0, 1. We use the standard complex
structure on $S^2 = {\Bbb C}~\cup~\{\infty\}$ to be covered by two
charts ${\Bbb C}$ and ${\Bbb C}^*~\cup~\{\infty\}$ with local complex
coordinates $z$ and $w$, respectively, related by the transition
function $w = -1/z$ in the overlap of the charts. After discussing the
action of the model we show that the classical dynamics of
($m,Y$)-superparticle is completely determined by the presence of two
first-class and $4N$ second-class constraints in the Hamilton
formalism. The first-class constraints strongly commute with all the
constraints and turn out to be classical counterparts for the Casimir
operator of the super-Poincar\'e group. We restrict ourselves, for the
sake of technical simplicity, by the case of $N$ = 1 supersymmetry in
considering covariant quantization of the model. Central-charges extensions
of the model are briefly discussed.

Let us begin with a statement that the supermanifold ${\Bbb R}^{4|4N}
\times S^2$ is a homogeneous space for the
($N$-extended)super-Poincar\'e group. An action of the four-dimensional
Lorentz group $SL(2,{\Bbb C})/{\Bbb Z}_2$ on $S^2$ is specified by
associating with a matrix
$$
({N_\alpha}^\beta) = \left( \begin{array}{cc} a & b\\
c & d \end{array} \right) \in SL(2,{\Bbb C}), \qquad \alpha,\beta = 0,1,
\eqno{(1)}$$
for any $N \in SL(2,{\Bbb C})$, the following complex automorphism of
$S^2$
$$
z \rightarrow z' = \frac{az-b}{-cz+d}.
\eqno{(2)}$$
The space-time translations and supersymmetry transformations are
defined to act trivially on $S^2$. Introducing two-component objects
$$
z^\alpha \equiv (z)^\alpha = (1,z), \qquad {\bar z}^{\dot\alpha} =
(z^\alpha)^*, \qquad \alpha, \dot\alpha = 0, 1,
\eqno{(3)}$$
one can rewrite Eq. (2) in the form
$$
z^\alpha \rightarrow {z'}^\alpha = \left(\frac{\partial z'}{\partial z}
\right)^{1/2} z^\beta {(N^{-1})_\beta}^\alpha .
\eqno{(4)}$$
This relation means that $z^\alpha$ is transforming simultaneously as a
left-handed Weyl spinor and a spinor tensor field on $S^2$. Owing to
Eq. (4), all the subsequent relations are explicitly Lorentz covariant.

\addtocounter{footnote}{-2}
The Lagrangian of ($m,Y$)-superparticle reads\footnote{Our
two-component spinor notation and convention mainly coincide with those
adopted in [2], except we number spinor indices by values 0, 1 to have
Eqs. (3), (4) and define spinning matrices $\sigma_{ab}$ and
${\tilde\sigma}_{ab}$ with additional minus sign in comparison with
[2].}
$$
{\cal L} = \frac{1}{2e_1}(\Pi^2 - (me_1)^2) + \frac{1}{2e_2}
\left(\frac{4{\dot z}{\dot{\bar z}}}{(\Pi^a\xi_a)^2} e_1^2 +
(\Delta e_2)^2 \right).
\eqno{(5)}$$
where
$$
\Pi^a = {\dot x}^a + {\rm i}\theta^I\sigma^a{\dot{\bar\theta}}_I -
{\rm i}{\dot\theta}^I\sigma^a{\bar\theta}_I;
\eqno{(6.a)}$$
$$
\xi_a = (\sigma_a)_{\alpha{\dot\alpha}} z^\alpha{\bar z}^{\dot\alpha},
\qquad \xi^a\xi_a \equiv 0;
\eqno{(6.b)}$$
$$
\Delta = m\sqrt{Y(Y+1)}.
\eqno{(6.c)}$$
Here $e_1(\tau)$ and $e_2(\tau)$ are Lagrange multipliers, for $\tau$
the evolution parameter, and the dots in Eqs. (5) and (6.a) mean, as usual,
${\rm d}/{\rm d}\tau$. The Lagrangian is seen to be manifestly
super-Poincar\'e invariant and remains unchanged under global $U(N)$
internal transformations. In addition, the system possesses two local
symmetries. Really, the action functional is obviously invariant under
arbitrary reparametrizations of the form
$$
\delta_\epsilon x^a = {\dot x}^a\epsilon, \qquad
\delta_\epsilon \theta^{\alpha I} = {\dot\theta}^{\alpha I}\epsilon,
\qquad \delta_\epsilon z = {\dot z}\epsilon ,
$$
{~}\hfill (7)\\
$$
\delta_\epsilon e_1 = \frac{{\rm d}}{{\rm d}\tau}(e_1\epsilon ),
\qquad \delta_\epsilon e_2 = \frac{{\rm d}}{{\rm d}\tau}(e_2\epsilon ).
$$
Another gauge symmetry looks like
$$
\delta_\mu x^a = p^a\mu , \qquad \delta_\mu \theta^{\alpha I} =
\delta_\mu z = 0,
$$
{~}\hfill (8)\\
$$
\delta_\mu e_1 = \dot\mu , \qquad \delta_\mu e_2 = 0,
$$
where $p^a$ is the canonically conjugate momentum to $x^a$,
$$
p_a = \frac{\partial{\cal L}}{\partial{\dot x}^a} = \frac{\Pi_a}{e_1} -
 \frac{4{\dot z}{\dot{\bar z}}}{(\Pi,\xi)^3}~ \frac{e_1^2}{e_2} \xi_a.
\eqno{(9)}$$

The multipliers $e_1$ and $e_2$ can be easily eliminated with the help
of their equations of motion. Then the Lagrangian (5) reduces to the
form
$$
\widehat{\cal L} = -mc\sqrt{-\Pi^2\left( 1 - \frac{4\Delta}{m^2c^2}~
\frac{|{\dot z}|}{(\Pi,\xi)}\right)},
\eqno{(10)}$$
what will be our starting point in passing to the Hamilton formulation
of the model.

It should be stressed that our consideration is restricted to the case
of supersymmetry without central charges. To include central charges,
one simply is to modify the Lagrangian (10) by adding a central-charges
term similarly to that it was proposed for the scalar superparticle by
Azcarraga and Lukierski [3]. There are two possible formulations, with
and without use of central-charges variables. The former consists in
introducing $N(N-1)$ new bosonic configuration-space variables
$\Lambda^{IJ} = -\Lambda^{JI}$ and ${\bar\Lambda}_{IJ}$ ($N$ even, for
$\bar\Lambda$'s the conjugates of $\Lambda$'s), and replacing the
Lagrangian (10) by
$$
\widehat{\cal L} + m\sqrt{N\Omega^{IJ}{\overline\Omega}_{IJ}} \vspace{-12pt},
$$
{~}\hfill \vspace{-12pt}(11)\\
$$
\Omega^{IJ} = {\dot\Lambda}^{IJ} + \theta^{[I}{\dot\theta}^{J]}.
$$
The latter is obtained by replacing $\widehat{\cal L}$ with
$$
{\widehat{\cal L}} + {\rm i}m(\theta^I C_{IJ}{\dot\theta}^J -
{\bar\theta}_I {\overline C}^{IJ}{\dot{\bar\theta}}_J),
\eqno{(12)}$$
$C$ being some constant antisymmetric matrix related to the central charges
(the original global $U(N)$ internal symmetry is broken here to $Sp(N/2)$,
$N$ even). As is known, the models of Ref. [3] and their generalizations have
aroused considerable interest last years (see, e.g., [4--6]). due to
the fact that they possess a certain type of fermionic local invariance
called Siegel symmetry [7] which is specific for the Brink--Schwarz
superparticle [8] and the Green--Schwarz superstring [9]. The models
(11) and (12) turn out to possess (for special $C_{IJ}$) such local
fermionic symmetries too, what we are planning to discuss in a separate
publication. It is worth remarking, however, that the gauge invariances
(7), (8) survive in both models (11) and (12).

Starting with the Lagrangian (10), we now introduce the momenta
$$
p_a = \widehat{\cal L}\frac{\stackrel{\leftarrow}{\partial}}{{\dot
q}^A} = (p_a, \pi_{\alpha I}, {\bar\pi}_{\dot\alpha}^{~I}, p_z, p_{\bar
z})
$$
conjugate to the coordinates $q^A = (x^a, \theta^{\alpha I},
{\bar\theta}^{\dot\alpha}_{~I}, z, {\bar z})$. One readily finds the
primary constraints
$$
T_1 \equiv p^2 + m^2 \approx 0,
\eqno{(13.a)}$$
$$
T_2 \equiv (p,\xi)^2p_zp_{\bar z} - m^2Y(Y+1) \approx 0,
\eqno{(13.b)}$$
$$
{\cal D}_{\alpha I} \equiv - {\rm i}\pi_{\alpha I} - p_{\alpha
\dot\alpha}{\bar\theta}^{\dot\alpha}_{~I} \approx 0,
\eqno{(13.c)}$$
$$
\left.{\overline{\cal D}}_{\dot\alpha}\right.^I \equiv  {\rm i}\left.
{\bar\pi}_{\dot\alpha}\right. ^I + p_{\alpha \dot\alpha}\theta^{\alpha I}
\approx 0,
\eqno{(13.d)}$$
where $p_{\alpha \dot\alpha} = (\sigma_a)_{\alpha \dot\alpha} p^a$.
These constraints constitute the full constrained surface in the phase
space, since the Hamiltonian of the system identically equals to zero,
as a consequence of reparametrization invariance. Thus the total
Hamiltonian reads
$$
H = {1\over2}e_1T_1 + {1\over2}e_2T_2 + \lambda^{\alpha I} {\cal
D}_{\alpha I} + \left.{\bar\lambda}^{\dot\alpha}\right._I
\left.{\overline{\cal D}}_{\dot\alpha}\right.^I,
\eqno{(14)}$$
where $e_1$, $e_2$, $\lambda^{\alpha I}$ and
${\bar\lambda}_{\dot\alpha I}$ are Lagrange multipliers. With respect
to the canonical graded Poisson bracket
$$
\{F,G\}_{\rm PB} = F\frac{\stackrel{\leftarrow}{\partial}}{\partial
q^A}~ \frac{\stackrel{\rightarrow}{\partial}}{\partial p_A}G -
(-1)^{\varepsilon(F)\varepsilon(G)}
G\frac{\stackrel{\leftarrow}{\partial}} {\partial
q^A}~ \frac{\stackrel{\rightarrow}{\partial}}{\partial p_A}F,
$$
for $F$ and $G$ phase-space functions having parities $\varepsilon(F)$
and $\varepsilon(G)$ respectively, the algebra of constraints is
$$
\{{\cal D}_{\alpha I}, \left.{\overline{\cal D}}_{\dot\alpha}\right.^J
\}_{\rm PB} = -2{\rm i}{\delta_I}^Jp_{\alpha\dot\alpha},
\eqno{(15)}$$
other brackets of constraints vanish strongly. Therefore, $T_1$ and
$T_2$ are the Abelian first-class constraints, while ${\cal D}_{\alpha
I}$ and $\left.{\overline{\cal D}}_{\dot\alpha}\right.^I$ are the second-class
constraints in accordance with Eq. (13.a). The second-class constraints
define the surface ${\cal D}_{\alpha I} = \left.{\overline{\cal
D}}_{\dot\alpha}\right.^J = 0$, or the reduced phase space, on which the
bracket $\{~,~\}_{\rm PB}$ is to be replaced by the Dirac one
$\{~,~\}_{\rm DB}$. The nonvanishing fundamental DB's are the following
$$
\{x^a,p_b\}_{\rm DB} = {\delta^a}_b, \qquad \{z,p_z\}_{\rm DB} = 1,
$$
$$
\{\theta^{\alpha I},{{\bar\theta}^{\dot\alpha}}_J\}_{\rm DB} = {{\rm
i}\over2}{p^{\alpha\dot\alpha}\over p^2}{\delta^I}_J, \qquad
\{\theta^{\alpha I},x^{\beta\dot\beta}\}_{\rm DB} =
{p^{\alpha\dot\beta}\over p^2} \theta^{\beta I},
\eqno{(16)}$$
$$
\{x^{\alpha\dot\alpha},x^{\beta\dot\beta}\}_{\rm DB} = {2{\rm i}\over
p^2}(\theta^{\alpha I}\left.{\bar\theta}^{\dot\beta}\right._I
p^{\beta\dot\alpha} - \theta^{\beta I}\left.{\bar\theta}^{\dot\alpha}
\right._I p^{\alpha\dot\beta}) ,
$$
and their conjugates.

It is interesting to note that the quantization via reduction procedure
has an alternative for the model under consideration, because the set
of second-class constraints ${\cal D}_{\alpha I},{\overline{\cal
D}}_{\dot\alpha}^{~I}$ is consistent with the split involution
requirements [10] by virtue of Eq. (15). Thus the model could be
BFV--BRST canonically quantized in terms of the full phase space on the
base of the method proposed in [10].

The constraints (13.a) and (13.c,d) are known to be standard for
superparticle models (the functions ${\cal D}_{\alpha I}$ and ${\overline
{\cal D}}_{\dot\alpha}^{~I}$ being phase-space analogs of the spinor
covariant derivatives), while Eq. (13.b) presents itself the specific feature
of our model. To understand its significance, let us study global
symmetries of the model in phase space, the full symmetry group being
the product of the super Poincar\'e group and $U(N)$ internal group. We
consider an infinitesimal super Poincar\'e transformations acting on
the configuration space by the law
$$
\delta x^a = b^a + {K^a}_bx^b + {\rm i}\theta^I\sigma^a{\bar\epsilon}_I
- {\rm i}\epsilon^I\sigma^a \vspace{-12pt} {\bar\theta}_I,
$$
{~}\hfill \vspace{-12pt}(17)\\
$$
\delta{\theta_\alpha}^I = {K_\alpha}^\beta{\theta_\beta}^I +
{\epsilon_\alpha}^I, \qquad \delta z = K_{\alpha\beta}z^\alpha z^\beta,
$$
where the rigid parameters $b, K$ and $\epsilon, \bar\epsilon$ generate
the translation, Lorentz and supersymmetry transformations
respectively, $K_{ab} = -K_{ba}$, ${K_\alpha}^\beta = {1\over2} K_{ab}
{(\sigma^{ab})_\alpha}^\beta$. This transformation acts canonically on
the phase space such that the associated automorphism in the algebra of
phase-space functions reads
$$
\delta F = \{F, -b^a{\cal P}_a + {1\over2}K^{ab}{\cal J}_{ab} +
\epsilon^{\alpha I}{\cal Q}_{\alpha I} - {\bar\epsilon}^{\dot\alpha}_
{~I}{\overline{\cal Q}}_{\dot\alpha}^{~I}\}_{\rm PB},
\eqno{(18)}$$
for any phase-space function $F$. The Hamilton generators of the
super-Poincar\'e group look like
$$
{\cal P}_a = p_a, \qquad {\cal J}_{ab} = x_ap_b - x_bp_a +
\vspace{-12pt}{\cal M}_{ab},
$$
{~}\hfill \vspace{-12pt}(19)\\
$$
{\cal Q}_{\alpha I} = \pi_{\alpha I} + {\rm i}p_{\alpha\dot\alpha}
{\bar\theta}^{\dot\alpha}_{~I}, \qquad {\overline{\cal Q}}_{\dot\alpha}
^{~I} = -{\bar\pi}_{\dot\alpha}^{~I} - {\rm i}p_{\alpha\dot\alpha}
\theta^{\alpha I},
$$
and the spinning part of ${\cal J}_{ab}$ has the form
$$
{\cal M}_{ab} = {\widehat{\cal M}}_{ab} + {\breve{\cal M}}_{ab},
\eqno{(20.a)}$$
$$
{\widehat{\cal M}}_{ab} = -(\sigma_{ab})_{\alpha\beta}z^\alpha z^\beta
p_z + ({\tilde\sigma}_{ab})_{\dot\alpha\dot\beta} {\bar
z}^{\dot\alpha} {\bar z}^{\dot\beta} p_{\bar z},
\eqno{(20.b)}$$
$$
{\breve{\cal M}}_{ab} = (\sigma_{ab})_{\alpha\beta}
\theta^{\alpha I} {\pi^\beta}_I -
({\tilde\sigma}_{ab})_{\dot\alpha\dot\beta}
{\bar\theta}^{\dot\alpha}_{~I} {\bar\pi}^{\dot\beta I},
\eqno{(20.c)}$$
where $\widehat{\cal M}$ and $\breve{\cal M}$ correspond to
the spherical and spinor degrees of freedom.

It is obvious that the functions $T_1$ and $T_2$ are super-Poincar\'e-
and $U(N)$-invariants, whereas ${\cal D}_{\alpha I}$ and ${\overline
{\cal D}}_{\dot\alpha}^{~I}$ possess linear homogeneous transformation
laws under the full symmetry group, in particular
$$
\{{\cal D}_{\alpha I},{\cal Q}_{\beta J}\}_{\rm PB} = \{{\cal D}_{\alpha
I},{\overline{\cal Q}}_{\dot\beta}^{~J}\}_{\rm PB} = 0.
\eqno{(21)}$$
As a result, each of the reduced phase space and the full constrained
surface moves onto itself by arbitrary transformations from the full
symmetry group. There is no loss of manifest covariance upon
restricting the dynamics to the reduced phase space.

The $N$-extended Poincar\'e superalgebra without central charges,
spanned by ${\Bbb P}_a$, ${\Bbb J}_{ab}$, ${\Bbb Q}_{\alpha I}$ and
${\overline{\Bbb Q}}_{\dot\alpha}^{~I}$, possesses two Casimir
operators (see, e.g., [11])
$$
{\Bbb C}_1 = {\Bbb P}^2,
\eqno{(22.a)}$$
$$
{\Bbb C}_2 = ({\Bbb L},{\Bbb P})^2 - {\Bbb L}^2{\Bbb P}^2,
\eqno{(22.b)}$$
where
$$
{\Bbb L}_a = {1\over2}\varepsilon_{abcd}{\Bbb J}^{bc}{\Bbb P}^d -
{1\over8}(\sigma_a)_{\alpha\dot\alpha}[{{\Bbb Q}^\alpha}_I,
{\overline{\Bbb Q}}^{\dot\alpha I}]
\eqno{(23)}$$
is the supersymmetric Pauli--Lubanski vector. Equation (22.b) defines
the superspin operator. Now, let us consider phase space counterparts
of the Casimir operators
$$
{\cal C}_1 = p^2,
\eqno{(24.a)}$$
$$
{\cal C}_2 = -p^2({\widehat{\cal W}}^2 + {1\over2}{\widehat{\cal W}}^a{\cal
D}_I\sigma_a{\overline{\cal D}}^I) + {1\over16}(p^a{\cal D}_I\sigma_a
{\overline{\cal D}}^I)^2 - {1\over8}p^2{{\cal D}^\alpha}_I {\cal
D}_{\alpha J}{\overline{\cal D}}^{\dot\alpha I}{\overline{\cal
D}}_{\dot\alpha}^{~J},
\eqno{(24.b)}$$
where
$$
{\widehat{\cal W}}_a = {1\over2}\varepsilon_{abcd}{\widehat{\cal
M}}^{bc}p^d
\eqno{(25.a)}$$
is the classical Pauli--Lubanski vector of ($m,s$)-particle [1],
$$
{\widehat{\cal W}}^2 = {\widehat{\cal W}}^a{\widehat{\cal W}}_a =
(p^a\xi_a)^2p_zp_{\bar z}.
\eqno{(25.b)}$$
Equations (24.b) and (25.b) show that the restriction of ${\cal C}_2$
to the reduced phase space reads
$$
\left.{\cal C}_2\right|_{{\cal D}={\overline{\cal D}}=0} =
-p^2(p,\xi)^2p_zp_{\bar z}.
\eqno{(26)}$$
Therefore, the set of constraints (13.a--d) implies that the classical
counterparts of the Casimir operators (22) strongly conserve on the
full constrained surface,
$$
{\cal C}_1 \approx -m^2, \qquad {\cal C}_2 \approx m^4Y(Y+1).
\eqno{(27)}$$

Now we are going to establish the general structure of physical
observables in the model. As is known, a function over the reduced
phase space is said to be a strong (weak) physical observable if it
strongly (weakly) commutes with the first-class constraints. In the
model of ($m,Y$) superparticle, each strong physical observable ${\cal
F}$,
$$
\{{\cal F}, T_i\}_{\rm DB} = 0, \qquad i = 1,2,
\eqno{(28)}$$
turns out to be a function of the super-Poincar\'e generators (19) only
$$
{\cal F} = \varphi({\cal P}_a, {\cal J}_{ab}, {\cal Q}_{\alpha I},
{\overline{\cal Q}}_{\dot\alpha}^{~I}).
\eqno{(29)}$$
This can be proved by direct examination of Eq. (28). Any weak
observable appears to be the sum of a strong observable and a linear
combination of constraints. The physical observables can possess, in
fact, arbitrary analytic dependence on $\theta$ and $\bar\theta$, since
in the reduced phase space we have
$$
\theta^{\alpha I} = -~{{\rm i}\over 2p^2}p^{\alpha\dot\alpha}{\overline
{\cal Q}}_{\dot\alpha}^{~I}, \qquad {\bar\theta}_{\dot\alpha}^{~I} = {{\rm
i}\over 2p^2}p^{\alpha\dot\alpha}{\cal Q}_{\alpha I}.
\eqno{(30)}$$
For the phase-space $U(N)$ generators, which are strong physical
observables of the model, one gets
$$
{A_I}^J = {{\rm i}\over2}({\pi^\alpha}_I{\theta_\alpha}^J -
{\bar\pi}^{\dot\alpha J}{\bar\theta}_{\dot\alpha I}) \approx
{1\over 4p^2}p^a{\cal Q}_I\sigma_a{\overline{\cal Q}}^J.
\eqno{(31)}$$

Owing to the general structure of physical observables described, the
problem of covariant quantization of the model is equivalent to
obtaining an explicit realization for the irreducible mass-$m$,
superspin-$Y$ representation of the Poincar\'e superalgebra. The
requirement of irreducibility, which is expressed as ${\Bbb C}_1 =
-m^2{\bu}$ and ${\Bbb C}_2 = m^4Y(Y+1){\bu}$, automatically
yields imposing the quantized first-class constraints on the physical
wave-functions. In the super wave-function approach, the quantization
implies such a representation to be realized in a space of tensor
superfields over the configuration space ${\Bbb R}^{4|4N} \times S^2$.

For simplicity, we describe the physical super wave-functions only for
$N$ = 1. Let $Y$ be an integer, $Y = 0,1,2,\ldots$~. It turns out that
the space of ($m,Y$)-superparticle wave-functions can be embedded into
the space of scalar superfields over ${\Bbb R}^{4|4N} \times S^2$. The
corresponding super-Poincar\'e generators (${\bf P}_a, {\bf J}_{ab},
{\bf Q}_\alpha, {\overline{\bf Q}}_{\dot\alpha}$) and the superspace
covariant derivatives ${\bf D}_\alpha, {\overline{\bf D}}_{\dot\alpha}$
follow from Eqs. (19), (20) and (13.c,d), respectively, if the ordinary
coordinate representation is used
$$
p_A = -{\rm i}(-)^A\frac{\stackrel{\rightarrow}{\partial}}{\partial
q^A},
\eqno{(32)}$$
where $(-)^A$ is equal to +1 for even $q^A$ and -1 for odd $q^A$. In
this representation, the superspin operator (22.b) takes the form (see,
e.g., [11])
$$
{\bf C}_2 = -{\bf P}^2({\widehat{\bf W}}^2 + {1\over4}{\widehat{\bf
W}}^{\alpha\dot\alpha}[{\bf D}_\alpha,{\overline{\bf D}}_{\dot\alpha}])
+ {1\over64}({\bf P}^{\alpha\dot\alpha}[{\bf D}_\alpha,{\overline{\bf
D}}_{\dot\alpha}])^2 + {1\over32}{\bf P}^2[{\bf D}^\alpha,
{\overline{\bf D}}^{\dot\alpha}][{\bf D}_\alpha,{\overline{\bf
D}}_{\dot\alpha}].
\eqno{(33)}$$
Here ${\widehat{\bf W}}_a$ and ${\widehat{\bf W}}^2$ are the results of
substituting (32) in the phase-space functions in Eqs. (25.a) and
(25.b), respectively, for ${\widehat{\cal M}}_{ab}$ given as in (20.b).
The operator ${\bf C}_2$ presents the quantum version of (24.b), which
corresponds to some special ordering of ${\bf D},{\overline{\bf D}}$
factors. To single out a physical subspace, we can require the physical
states to be annihilated by the quantum first-class constraints,
$$
({\bf P}^2 + m^2)\Phi = 0,
\eqno{(34.a)}$$
$$
({\widehat{\bf W}}^2 - m^2Y(Y+1))\Phi = 0,
\eqno{(34.b)}$$
as well as by a half of the second-class ones. Upon fulfilment of Eq.
(34.a), Eq. (34.b) will imply the relation
$$
({\bf C}_2 - m^2Y(Y+1))\Phi = 0
\eqno{(35)}$$
if and only if the physical states are chosen to be chiral, ${\overline
{\bf D}}_{\dot\alpha}\Phi$ = 0, or antichiral, ${\bf D}_\alpha\Phi$ =
0. In the former case, the physical wave-functions are subjected to the
equations:
$$
\Box\Phi = m^2\Phi,
\eqno{(36.a)}$$
$$
(\xi^a\partial_a)^2\partial_z\partial_{\bar z}\Phi = m^2Y(Y+1))\Phi,
\eqno{(36.b)}$$
$$
{\overline {\bf D}}_{\dot\alpha}\Phi = 0.
\eqno{(36.c)}$$

The general solution to Eqs. (36.a--c) reads on the massive hyperboloid
in momentum space as
$$
\Phi(p,\theta,\bar\theta,z,{\bar z}) = \exp (-p^a\theta\sigma_a
{\bar\theta})\phi_{a_1\ldots a_Y}(p,\theta) \frac{\xi^{a_1}\ldots
\xi^{a_Y}}{(p,\xi)^Y},
\eqno{(37)}$$
where $\phi_{a_1\ldots a_Y}(p,\theta)$ is totally symmetric in its
indices, traceless and $p$-transversal:
$$
\phi_{a_1\ldots a_Y} = \phi_{(a_1\ldots a_Y)}, \qquad
{\phi^b}_{ba_1\ldots a_{Y-2}} = p^b\phi_{ba_1\ldots a_{Y-1}} = 0,
\eqno{(38)}$$
see Ref. [1] for more detail. Together with the on-shell condition $p^2 +
m^2 = 0$, these equations define an on-shell massive chiral superfield of
superspin $Y$ [11]. To have a real representation, Eq. (36.a) is to be
replaced by the stronger one
$$
-~{1\over4}{\bf D}^2\Phi = m{\overline\Phi}.
\eqno{(39)}$$

For a half-integer superspin, $Y = 1/2, 3/2, \ldots$, the space of
physical super wave-functions can be embedded into a space of superfields
over ${\Bbb R}^{4|4N} \times S^2$ transforming by the law
$$
\Psi'(q'^A) = \left(\frac{\partial z'}{\partial z}\right)^{1/2}
\Psi(q^A)
\eqno{(40)}$$
with respect to the super-Poincar\'e group, see Ref. [1] for more detail.

Now, let us summarize the results. In the present paper, we have constructed
the model for a free massive $N$-extended superparticle with any fixed
superspin, and considered its classical and quantum aspects. This theory
is the minimal supersymmetric generalization of the ($m,s$)-particle model
[1] previously developed. The non-minimal generalizations, which where outlined
above and shown to be consistent with the inherent gauge structure, correspond
to the case of supersymmetry with central charges. The model of
($m,Y$)-superparticle appears to admit a natural extension to anti-de Sitter
superspace [12]. Thus, our model gives a unified framework for descibing the
dynamics of arbitrary superspin massive superparticles.

\bigskip

\centerline{\bf Acknowlegments}}

\bigskip

This work was supported in part by the International Science
Foundation under the Emergency Grants Program.

\bigskip

\centerline{References}

\bigskip

\noindent
[1] S.M. Kuzenko, S.L. Lyakhovich, and A.Yu. Segal, Preprint
TSU-TP-94-8, hep-th/94 03196.\newline
[2] J. Wess and J. Bagger, {\it Supersymmetry and Supergravity}
(Princeton Univ. Press., Princeton, 1983).\newline
[3] J.A. Azc\'arraga and J. Lukierski, Phys. Lett. {\bf B113} (1982) 170;
Phys. Rev. {\bf D28} (1983) 1337.\newline
[4] P.K. Townsend, Phys. Lett. {\bf B202} (1988) 53.\newline
[5] J.A. Azc\'arraga and J. Lukierski, Phys. Rev. {\bf D38} (1988)
509.\newline
[6] J.M. Evans, Nucl. Phys. {\bf B331} (1990) 711.\newline
[7] W. Siegel, Phys. Lett. {\bf B128} (1983) 397.\newline
[8] L. Brink and J.H. Schwarz, Phys. Lett. {\bf B100} (1981) 310.\newline
[9] M.B. Green and J.H. Schwarz, Phys. Lett. {\bf B136} (1984) 357.\newline
[10] I.A. Batalin, S.L. Lyakhovich, and I.V. Tyutin, Mod. Phys. Lett.
{\bf A7} (1992) 1931.\newline
[11] S.J. Gates, M.T. Grisaru, M. Ro\v{c}ek, and W. Siegel, {\it
Superspace} (Benjamin--Cummings, Reading, MA, 1983).\newline
[12] S.M. Kuzenko, S.L. Lyakhovich, A.Yu. Segal, and A.A.
Sharapov, in preparation.
\end{document}